\documentclass[12pt,a4paper]{article}
\usepackage{graphics}
\usepackage{amsfonts}

\newcommand{\be}{\begin{equation}}
\newcommand{\ee}{\end{equation}}
\newcommand{\eel}[1]{\label{eq:#1}\end{equation}}
\newcommand{\bea}{\begin{eqnarray}}
\newcommand{\eea}{\end{eqnarray}}
\newcommand{\eeal}[1]{\label{eq:#1}\end{eqnarray}}
\newcommand{\baq}{\begin{equation}\begin{array}{rcl}}
\newcommand{\eaq}{\end{aryray}\end{equation}}
\newcommand{\eaql}[1]{\end{array}\label{eq:#1}\end{equation}}
\newcommand{\beac}{\begin{equation}\begin{array}{rcl}}
\newcommand{\eeacn}[1]{\end{array}\label{eq:#1}\end{equation}}
\newcommand{\ba}{\begin{array}}
\newcommand{\ea}{\end{array}}
\newcommand{\non}{\nonumber \\}
\newcommand{\equ}[1]{(\ref{eq:#1})}

\renewcommand{\a}{\alpha}

\newcommand{\notR}{\not{\hbox{\kern-4pt $R$}}}
\newcommand{\journal}[4]{{\rm #1~}{#2}\,(19#3)\,#4}

\newcommand{\np}{\journal {Nucl. Phys.}}
\newcommand{\pl}{\journal {Phys. Lett.}}

\begin{document}

\begin{titlepage}

\begin{flushright}
\textsf{TAUP--2427--97, LMU--TPW--97--15}\\
\textsf{hep--th/9705232}\\
\textsf{May 1997}
\end{flushright}
\vfill
\vspace{-1cm}
\begin{center}
\textbf{\large M Theory and Seiberg-Witten Curves: Orthogonal and
  Symplectic Groups}
\vskip 1cm
\textsc{A. Brandhuber$^a$, J. Sonnenschein$^{a,}$\footnote{Work supported
    in part by the US-Israel Binational Science Foundation, by GIF --
    the German-Israeli Foundation for Scientific Research, by the
    Israel Science Foundation
    and by the European Commission TMR programme ERBFMRX-CT96-0045, 
    in which S.T. is associated to HU-Berlin.},}\\
\textsc{S. Theisen$^{b,1}$ and S. Yankielowicz$^{a,1}$}\\[10mm]
\emph{$^a$School of Physics and Astronomy,}\\
\emph{Beverly and Raymond-Sackler Faculty of Exact Sciences,}\\
\emph{Tel-Aviv University, Ramat-Aviv, Tel-Aviv 69978, Israel}\\[5mm]
\emph{and}\\[5mm]
\emph{$^b$Sektion Physik, Universit\"at M\"unchen, }\\
\emph{Theresienstra\ss e 37, 80333 M\"unchen, FRG}

\end{center}
\vfill

\thispagestyle{empty}

\begin{abstract}

We discuss $N=2$ supersymmetric Type IIA brane configurations within
M theory. This is a generalization of the work of Witten to all classical 
groups. 

\end{abstract}

\end{titlepage}

\section{Introduction}

Our understanding of supersymmetric field theories was greatly advanced by 
the seminal work of Seiberg and Witten \cite{sw}. 
Very early on it has been suggested that 
there should also be important ramifications for string theory which have 
then, in due course, been found and worked out \cite{kklmv,kkv,bjpsv}.
The relation between 
string theory and supersymmetric field theory has become most transparent in
the work of Hanany and Witten \cite{han}
who turned the brane technology into 
an efficient and easy tool to engineer supersymmetric field theories. Various 
dualities could be demonstrated this way; they are very natural in the
brane picture \cite{egk,esj,bsty,bh,egkrs,tatar}; see also \cite{bjpsv,ov,ahn}
for an alternative brane picture of field theory dualities. 

One of the great surprises of the original work of Seiberg and Witten
was that the low energy effective action of asymptotically free N=2
supersymmetric field theories could be exactly computed. In this
computation, which is possible due to the holomorphic structure of the
Lagrangian, an auxiliary Riemann surface appears, whose period matrix
is identified with the gauge couplings of the theory in its Coulomb
phase. It was subsequently shown how this Riemann surface appears
geometrically in string compactification on Calabi-Yau manifolds as a
supersymmetric two-cycle around which the Type IIA twobrane wraps,
leaving a supersymmetric point-particle in uncompactified space-time 
\cite{kklmv,klmvw,lmw}; for excellent recent reviews, see 
\cite{lerche,klemm}.

Recently Witten \cite{witten}
has shown, how the Riemann surfaces naturally appear  
as supersymmetric cycles in the M theory context, by reading the N=2
supersymmetric brane configurations on the Type IIA theory as one convoluted 
M theory fivebrane, whose internal part, which extends into the
eleventh dimension, is a Riemann surface, holomorphically embedded into 
$\mathbb{R}^3\times\mathbf{S}^1$.
$\mathbb{R}^3$ are the three internal dimensions tangential to the
configuration of Dirichlet fourbranes (D4 branes) and NS fivebranes,
and $\mathbf{S}^1$ is the circle on which the eleventh dimension is
compactified. 
Independently, Evans, Johnson and Shapere \cite{esj}
also noticed the connection between M theory, Type IIA brane
configurations and Seiberg-Witten curves; their starting point was the
generalization of the work of \cite{egk} to orthogonal and symplectic
groups, with the orientifold plane playing a crucial role.
Witten's analysis was restricted to the case of $A_r$ gauge groups. We
present here the extension of his results to the other classical
groups. 

Orthogonal and symplectic groups have previously been discussed in the
brane context \cite{esj,egkrs,ov,ahn}. Here the appearance of an
orientifold plane complicates the discussion. Our goal is to
understand these theories in M theory. One line of attack would be to
argue for the brane configuration, which must of course respect the
orientifold symmetries, and then write down the equation for the
Riemann surface which should then agree with the hyperelliptic curves
which have been constructed purely from field theory
considerations. We have however chosen the reverse strategy, namely
starting from the known curves, we infer the brane
configurations. Here the curves which have been discussed in relation
of Seiberg-Witten theory with integrable systems are the most
appropriate. We find that the curve displays the orientifold plane
only indirectly, namely via the symmetry of the brane
configuration. Since in M theory the brane configuration is smooth,
there is no rationale for the jump in RR charge of the orientifold
fourplane (O4 plane) as it crosses a NS fivebrane. This was necessary in 
\cite{esj}
to explain the symplectic flavor symmetry of the dual orthogonal
gauge theory. In addition, in order to get a smooth transition from
the electric to the magnetic theory, the authors of ref.\cite{egkrs}
had to assume that two of the D4 branes which extend between the two
NS fivebranes, must position themselves on the O4 plane. In the M theory
picture we find two infinite D4 branes which account for both of these
phenomena. 

An outline of this paper is as follows. In section two we present our
interpretation of the Seiberg-Witten curves in M and Type IIA
theory. We also comment on the interpretation of the orientifold
plane. In this section matter always comes, in Type II language, from
D4 branes. D6 branes enter the stage in section 3. We end with
conclusions and an outlook. 

While we were completing this manuscript, a preprint by
K. Landsteiner, E. Lopez and D. Lowe, ``N=2 Supersymmetric Gauge
Theories, Branes and Orientifolds", hep-th/9705199 appeared, which
has substantial overlap with our work.

\section{Models With SO And Sp Gauge Groups}

\subsection{Generalities}

We will use the same conventions as \cite{witten}. The classical brane 
configuration in the Type IIA theory consists of infinite solitonic fivebranes
with worldvolume extending in the $x^0, x^1, x^2, x^3, x^4, x^5$
directions and Dirichlet fourbranes with worldvolume along $x^0, x^1, x^2,
x^3, x^6$. 
In section 3 we will also introduce D6 branes with worldvolume along 
$(x^0,x^1,x^2,x^3,x^7,x^8,x^9)$. All brane configurations considered preserve 
$1/4$ of the 32 supercharges of the Type IIA theory. As in
ref.\cite{witten} we have effectively a four-dimensional $N=2$
supersymmetric theory on the world volume of the D4 branes. 
We define 
\bea
v & = & x^4 + i x^5 \\
s & = & (x^6 + i x^{10})/R
\eea
where $x^{10}$ is a periodic coordinate $x^{10} \sim x^{10} + 2\pi R$. 
It is convenient to make a transformation from the cylinder with
coordinate $s$ to the complex plane $t = \exp{(-s)}$. 

In M theory (compactified in the $x^{10}$ direction on a circle of
radius $R$) the configuration is described by a single
fivebrane with a complicated worldvolume history
\be
\mathbb{R}^{3,1} \times \Sigma ~.
\ee
Here $\Sigma$ is a (non-compact) Riemann surface 
holomorphically embedded in the complex twoplane
parametrized by $v$ and $t$. $\Sigma$ is a supersymmetric cycle in the
sense of \cite{bbs}. 
Since in the Type IIA theory we deal with infinitely extended branes, 
the Riemann surface which appears in M-theory is non-compact. 
Here we always refer to the compactified surface; see
\cite{witten} for explanation.
We will now discuss the case of single $B_r,\,C_r$ and $D_r$ group
factors, followed by the discussion of multiple group factors. The
starting point will be the spectral curves associated with the various
simple groups, as given in \cite{mw,mmm}.  

\subsection{SO(2r)}

For the curve for N=2 gauge theories with gauge
group $SO(2r)$  we take
\bea
F(t,v)  = t^2 v^2 + 2 t P_r (v) + v^2 = 0
\eeal{curve}
with 
\be
P_r(v) = v^{2r} + c_2 v^{2r-2} + c_4 v^{2r-4} + \ldots + \tilde{c}_r^2 ~.
\ee
$c_{2n}\,,~n=1 \ldots r-1$ denote gauge invariant operators
(Casimirs) of order $2n$ and $\tilde{c}_r$ is the exceptional Casimir
of order $r$. Via the substitution $t \to (t-P_r)/v^2$ these curves
assume the form of the $SO(2r)$ hyperelliptic curves of
ref. \cite{bl}.  Substituting $t\to t/v^2$ we find the $D_n$ curves of
refs.\cite{mw,mmm}. 

For fixed $v$ the polynomial is of degree two in $t$ and the two roots
of eq. \equ{curve} correspond to the fact that we have 
a Type IIA configuration with two NS fivebranes. In general the
degree of the polynomial in $t$ equals the number of NS fivebranes.

On the other hand $F(t,v)$ is of degree $2r$ in $v$, reflecting the
presence of 
$2r$ D4 branes. Since the polynomial is even in $v$ the
configuration is symmetric under $v \to -v$. This hints towards the
existence of an orientifold plane at  $v=0$, parallel to the D4 branes, in the
classical Type IIA picture. This O4 plane enforces a reflection 
symmetric brane configuration. This incidentally, automatically removes the 
IR divergence in the five-brane kinetic energy, which, in the $U(r)$
case discussed in \cite{witten} led to a freezing out of the
$U(1)\subset U(r)$ factor associated with the motion of the center of
position of the D4 branes. 

To get further information on the brane
configuration, we now investigate the behavior of the curve in certain
limits. 

First we want to look at the limit where $v$ is
small. If $\tilde{c}_r = 0$ the polynomial factors into 
$v^2$ times a factor which is appropriate for the curve for gauge
group $SU(2r-2)$ with all odd Casimir invariants set to zero. This
corresponds to the situation where two infinite D4 branes coincide at
$v=0$ which, naively, would imply additional massless states from zero
length strings between these two branes. But from
a careful analysis in field theory \cite{bl} we know that the
monodromy at this singularity is trivial. Therefore there are no
additional particles becoming massless.
This observation was first made in the context of Type IIA brane 
configurations in ref.\cite{esj}.

In the case of
non-zero $\tilde{c}_r$ the curve becomes, again in the limit of small $v$, 
\be
t v^2 + 2 \tilde{c}_r^2 + v^2/t = 0~.
\ee
Going to small  $t$ (i.e. $s \to \infty$) requires $t\sim v^2$, 
which means that two roots of $F(v)$ asymptotically
approach $v = 0$. For large $t$ ($s \to -\infty$) we find $t\sim1/v^2$
which indicates two roots of $F(v)$ approaching $v=0$. 
We interpret this as two infinite D4 branes 
which are deformed in the region of small $x^6$ but approach 
the position $v=0$ as $x^6 \to\pm \infty$.

The question of how to identify segments of the curve with branes of Type IIA
can be addressed as follows. Equation \equ{curve} defines a multivalued map
from the $v$ plane to the $t$ plane with $4r$ branch points. 
We want to identify the objects that extend to $x^6\to\infty$ with $v$ small.
Examining the map mentioned above, we find that circling $v=0$ once
maps to circling $t=0$ {\it twice}. But 
a closed contour around the origin of the $t$ plane means going around
the $\mathbf{S}^1$. A D4-brane is distinguishable from a NS brane in
that the former wraps $\mathbf{S}^1$ and the latter does not. We thus
identify the objects that stretch to $x_6=\pm \infty$ as {\it two} D4 brane. 
Another way to see that we are dealing with two semi-infinite D4
branes is that for the codimension one subspace of the moduli space
defined by $\tilde c_r=0$ we have two branes precisely at $v=0$, which
means that they have to wrap around the $\mathbf{S}^1$. By continuity
this also holds for generic values of $\tilde c_r$. 

Let us now consider the situation where $v$ is large. 
For $t$ large ($s \to -\infty$), the roots for $v$ are approximately at
\be
t \equiv v^{2r-2} ~,
\ee
and for $t$ very small ($s \to \infty$) approximately at
\be
t \equiv v^{-(2r-2)} ~.
\ee
This describes the bending of an NS fivebrane when a net number of
$2r-2$ D4 branes end on it from the right and the left,
respectively. Again, note that the D4 branes are located symmetrically
with respect to the $v=0$ plane. $2(r-1)$ of the D4
branes have the same asymptotics which diverges exponentially. It is
to be identified with the position of the NS fivebranes at large $v$. 
In addition there are two infinite D4 branes which
asymptotically approach $v = 0$, the position of the IIA orientifold
plane which is not directly visible in the M theory picture; it is 
encoded in the curve only through the symmetry and the presence of the 
infinite D4 branes. 

As an example we have drawn in figure 1 the M5 brane for gauge group
$SO(10)$ with generic values of the moduli. Each line corresponds to
two D4 branes due to the $v \to -v$ symmetry.  More specifically we
have chosen a specific slice through $\Sigma$ with $x^{10} =
0$. Therefore, each of the lines is actually a tube (times
$\mathbb{R}^{3,1}$). The horizontal axis corresponds to $x^6$ and the
vertical axis to the absolute value of $v$. 

\begin{figure}
\begin{center}
 \resizebox{8cm}{!}{
   \includegraphics{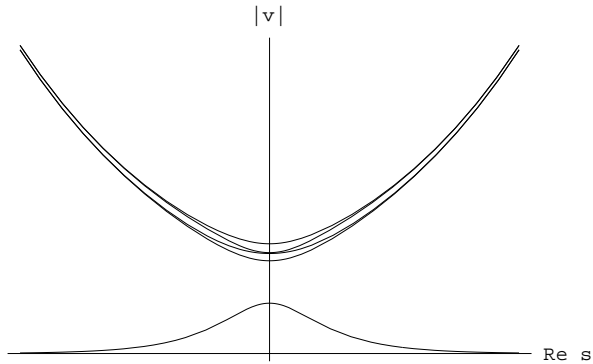}
   }
\end{center}
\caption{The M5 brane for gauge group $SO(10)$ }
\end{figure}

The generalization to the cases with $N_f=N_1+N_2$ matter multiplets is
\be
v^2 t^2\prod_{j=1}^{N_1}(v^2-m_j^2) + 2 t P_r(v)  + v^2 \prod_{i=N_1
  + 1}^{N_f} (v^2 - m_i^2) =0~,
\ee
which corresponds to adding $N_1$ semi-infinite mirror pairs of
fourbranes to the left of all fivebranes  and $N_2$ to the right. 

\subsection{SO(2r+1)}

\begin{figure}
\begin{center}
 \resizebox{8cm}{!}{
   \includegraphics{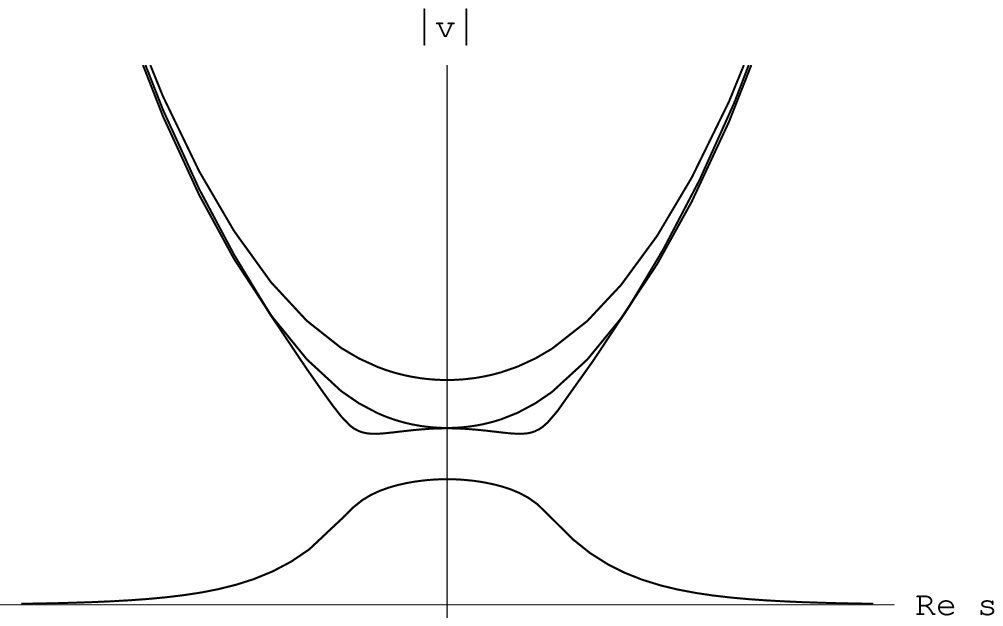}
   }
\end{center}
\caption{$SO(5)$}
\end{figure}

For gauge groups $SO(2r+1)$ we need a brane setup with $2r+1$ D4 branes
between the two NS fivebranes. This is achieved by curves of the form
\be
F(t,v) = t^2 v^2 + 2 t v P_r (v^2) + v^2 = 0 ~,
\quad
P_r(v) = v^{2r} + c_2 v^{2r-2} + c_4 v^{2r-4} + \ldots + c_{2r} ~.
\ee
which are related, via the
substitution $t\to t/v^2$ to the  $B_r$ curves of
\cite{mw,mmm}, whereas $t \to (t-P)/v$ reproduces those of \cite{ds}.
Important is the overall factor $v$, which is needed to get the 
correct number of D4 branes. 
Note that the same argument that was used in section 2.2 to 
demonstrate that there are indeed two semi-infinite D4 branes stretching
to $x^6\to\pm\infty$, now tells us that there is just one of these branes. 
In addition there is still the infinite D4 brane at $v=0$ which necessarily 
wraps around the $\mathbf{S}^1$.

Figure 2 shows an example with gauge group $SO(5)$. Every branch
corresponds now to one D4 brane in contrast to the $SO(2r)$ case because of
a different reflection symmetry; cf. below. 
Furthermore, the horizontal line
corresponds to the additional infinite D4 brane at $v=0$.

We now have again two D4 brane in the region to the left and to the 
right of the NS branes. 
As for $SO(2r)$ they approach $v=0$ as $x^6 \to \pm
\infty$. $F(t,v)=0$ is now invariant under $(t,v) \to (-t,-v)$. Note
that $t\to -t$ implies $x^{10}\to x^{10}+ \pi R$. This means that the
orientifolding also involves a non-trivial transformation in the
$x^{10}$ direction.  

In the $SO(2r+1)$ case the singularity at $c_{2r}=0$ has a non-trivial
monodromy, in contrast to the $\tilde{c}_r = 0$ singularity of the
$SO(2r)$ case, and a dyon becomes massless which is related to a short
root of the Lie algebra of $SO(2r+1)$ \cite{bl}. 
In the Type IIA picture there is an additional
D4 brane on top of the orientifold plane such that there are
additional states from strings between this special D4 and the other D4
branes. These states are necessary to lift the $SO(2r)$ multiplet to an
$SO(2r+1)$ multiplet. In our M theory configuration we have taken into
account this additional D4 at $v=0$ by introducing an additional
factor of $v$ in the curve. As $c_{2r} \to 0$ the curve develops
another infinite D4 brane on top of that which gives rise to an additional
massless state, as expected. 

Adding matter in the fundamental representation is straightforward: we
attach $N_1$ semi-infinite D4 branes (and their mirror images) from the left
to the left NS 5 brane and $N_2$ semi-infinite D4 branes (and their
mirror images) from the right to the right NS 5 brane, with
$N_1+N_2=N_F$ being the total number of fundamental flavors. The curve
then takes the form
\be
v^2 t^2 \prod_{j=1}^{N_1}(v^2-m_j^2)  + 2 t v P_r(v)  + v^2\prod_{i=N_1
  + 1}^{N_f} (v^2 - m_i^2) =0~.
\ee

\subsection{Sp(2r)}

We take the holomorphic curve which is to represent the brane
configuration in the pure gauge case to be 
\be 
t^2+2 t v^2 P_r(v) +1=0\,,~~P_r (v)=v^{2r}+c_2 v^{2r-2}+\dots+c_{2r}\,.
\ee
Note the symmetry under $v\to -v$. Via the substitution $t\to t v^2$
this curve is seen to be equivalent to the $C_r$ curve in \cite{mmm}.
The relation with the curves of \cite{as} is however not
so obvious. \footnote{We acknowledge instructive correspondence about
the various forms of the curves with A. Mironov and A. Morozov. Note
that there is also a disagreement between the $C_r$ curves of
\cite{mw} and \cite{mmm}.}

Figure 3 shows the generic form of the curve for gauge group
$Sp(6)$. As in the $SO(2r)$ case each branch corresponds to two D4
branes. From the form of the curve we read off that there are two
additional D4 branes between the two NS fivebranes. 
\begin{figure}
\begin{center}
 \resizebox{8cm}{!}{
   \includegraphics{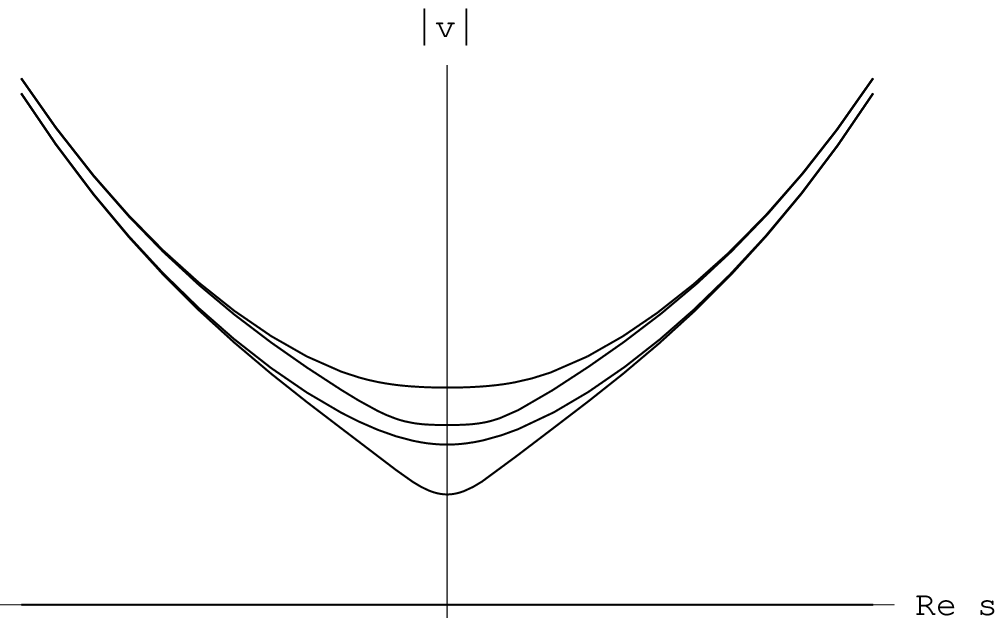}
   }
\end{center}
\caption{$Sp(6)$} 
\end{figure}
These two additional D4 branes have the property that they touch the
orientifold at $v=0, t_{1,2} = \pm i$ for all values of the
moduli. This is not visible in  picture 3 since we took a slice where
$t$ is real. 

Again, adding matter in the fundamental $2r$--dimensional representation 
of $Sp(2r)$ is straightforward: we simply add $N_1$ and  $N_2$
semi-infinite D4 branes (and their mirror images) to the left and to
the right, respectively, with $N_1+N_2=N_F$. 

We have thus seen that there is a natural correspondence between the
$A_r,B_r,C_r,D_r$ Seiberg-Witten curves as spectral curves of
appropriate integrable systems and the Type IIA brane configuration
which leads to $SU(r)$, $SO(2r)$, $SO(2r+1)$ and $Sp(2r)$ gauge theories
on the D4 branes. In the M theory context these curves simply describe
the internal part of the five-brane.

\subsection{The Orientifold Plane}

We would now like to examine the question how the structure of an
orientifold emerges from the M theory curves that correspond to the
orthogonal and symplectic groups.
In string theory we have, as a consequence of dividing by the
world sheet parity inversion times a space-time symmetry, an
orientifold plane which also carries RR charge, namely if $p$ is the
dimension of the orientifold plane, $\pm 2^{p-5}$ units of charge of a
physical Dp brane, i.e. the Dp brane and its mirror image. 
If we normalize the RR charge of a physical D4 brane
to be +2, the charge of an O4 plane is $\pm1$. In the geometric brane
arrangement of the Type IIA theory, the orientifold plane makes its
appearance by enforcing symmetry under reflection on the orientifold
plane, but, in the discussion of Seiberg dualities with SO and Sp
gauge groups, also via its RR charge, in particular via its charge
induced on the NS branes. Central in the discussion of \cite{esj} and
\cite{egkrs} was the fact that the charge of the orientifold plane
switches sign on traversing a NS fivebrane, so that in the simplest
arrangement of two NS fivebranes its charge is $-1$  between the five
branes and +1 outside. This is for orthogonal gauge groups and 
the sign of the charge is reversed for symplectic groups. 

Let us now see how the
orientifold plane and these charge assignments might be understood
from the M-theoretic point of view. We saw that with the $SO(2r)$
projection we get additional semi-infinite D4 branes, two on each side
of the  NS fivebrane arrangement, while with the $SO(2r+1)$ projection we
get one infinite D4 and one additional semi-infinite D4 brane on each
side. The $Sp$ projection leads to two
additional D4 branes between the two NS fivebranes. If we now agree to
assign RR charge {$-1$} to an infinite four-dimensional four-plane along
the $(x^1,x^2,x^3,x^6)$ direction, with $x^4 = x^5 = 0$ and $x^7,x^8,x^9$
fixed by the NS fivebranes, we find that the $SO(2r)$ projection effectively
leads to a charge assignment $(+1,-1,+1)$ for the three regions to the
left, between and to the right of the two NS fivebranes. For the $Sp$
projection we get $(-1,+1,-1)$ and for the $SO(2r+1)$ projection
$(+1,0,+1)$ instead.

\subsection{Product Gauge Groups}

We now consider more general models with chains of NS
fivebranes connected by D4 branes. The $n+1$ fivebranes are labeled
from $0$ to $n$ and the $(\a - 1)$-th fivebrane is connected to the $\a$-th 
fivebrane by $k_\a$ D4 branes.

As in the $N = 1$ supersymmetric case \cite{tatar} it is not possible
to create models with gauge groups $SO-SO-\ldots$ or $Sp-Sp-\ldots$, only
alternating chains of the form $SO-Sp-SO- \ldots$ or $Sp-SO-Sp-
\ldots$ are possible.  E.g. The second possibility is realized by the
following curve:
\be
F(t,v) = t^n + t^{n-1} v^2 P_1(v) + t^{n-2} P_2(v) + t^{n-3} v^2
P_3(v) + t^{n-4} P_4 + \ldots = 0
\ee
with
\be
P_\a(v) = v^{2 r_\a} + c^{(\a)}_2 v^{2 r_\a-2} + c^{(\a)}_4 v^{2 r_\a-4} +
\ldots +  c^{(\a)}_{2 r_\a} ~.
\ee
{}From the previous sections it is clear that this must correspond to a
gauge theory with gauge group
\be
G = Sp(2 r_1) \otimes SO(2 r_2) \otimes Sp(2 r_3) \otimes \ldots
\ee
and matter content
\be
\bigoplus_{\a=1}^{n-1} (2 r_\a,2 r_{\a+1}) ~.
\ee
In the case that the chain starts with an $SO$ gauge factor the first
terms of the curve are 
\be
t^n v^2 + t^{n-1} P_1(v) + t^{n-2} v^2 P_2(v) + \ldots = 0 ~.
\ee
In Figure 4 we have drawn a simple example with three gauge group
factors $G = Sp(6) \times SO(10) \times Sp(6)$. The branches going to
infinity for large Re $s$ correspond to the four NS fivebranes. They
are all bent differently according to the number of D4 branes ending
on them from the left and the right. Each branch corresponds to two D4
branes in a Type IIA configuration.
\begin{figure}
\begin{center}
 \resizebox{8cm}{!}{
   \includegraphics{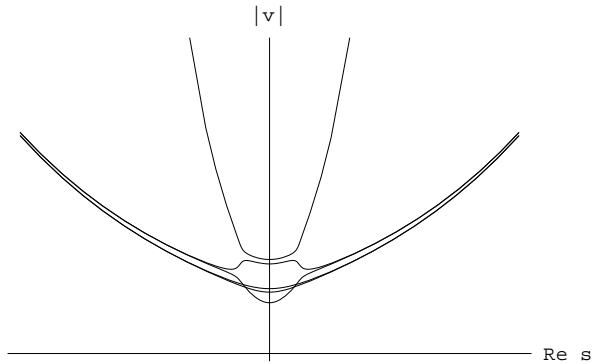}
   }
\end{center}
\caption{$Sp(6) \times SO(10) \times Sp(6)$}
\end{figure}
Chains without matter and $SO(2r+1)$ factors are not possible as 
follows immediately from our discussion of the curves and the brane 
configurations derived from it. 

We can add matter by putting semi-infinite D4 branes in the usual
way. For example we can multiply 
the highest power of $t$ by $\prod_{i=1}^{N_f} (v^2 - m_i^2)$
which will lead to $N_f$ matter hypermultiplets in the
fundamental representation of the first gauge group factor, $SO(2
r_1)$ or $Sp(2 r_1)$. 

If we compactify these chains on a circle $\mathbf{S}^1$ in the $x^6$
direction we find the generalizations of the elliptic models of
\cite{witten}. This means that the semi-infinite D4 branes to the left
and to the right are connected and produce an additional gauge group
factor. This is only consistent if we have an even number of NS branes. 
In this case the gauge group is
\be
G = SO(2 r_1) \times Sp(2 r_2) \times SO(2 r_3) \times Sp(2 r_4)
\times \ldots \times  Sp(2 r_{2 n}) ,
\ee
and matter comes in the usual mixed representations with the exception
that there is an additional hypermultiplet in the $(2 r_{2n}, 2 r_1)$
representation. If the number of D6 branes $b_l = 0$ (cf. below)
one can make the
beta functions vanish for all gauge group factors.
\be
r_{2 i +1} = r_{2 j} + 1 = n + 1, \textrm{for} ~1 \leq i,j \leq n~.
\ee
In this case the gauge group is $SO(2 n + 2) \times Sp(2n) \times \ldots$.

\section{Configurations With D6 Branes}

We now incorporate D6 branes in our 
configurations, following closely ref.\cite{witten}.
We now place $d_\a$ D6 branes between the
$(\a -1)$-th and $\a$-th NS fivebrane. Each D6 brane is located at
definite values of $x^4$,$x^5$ and $x^6$. Due to the symmetry $v \to-v$ 
we always have to place D6 branes in pairs at locations $v$ and $-v$. 

The interpretation of the resulting world-volume theory is clear with
gauge group and matter from D4 branes as before.
In addition we have $d_\a$ hypermultiplets in the fundamental
representation of the corresponding orthogonal or symplectic gauge
group. The $v$ positions of the D6 branes give the bare
masses. Their $x^6$ positions decouple from the low energy
four-dimensional physics; they become relevant in the discussion of
Higgs or mixed branches only \cite{witten}. 

In this section we will discard the semi-infinite D4 branes  
which gave rise to (massive) hypermultiplets; we can
generate an arbitrary number of hypermultiplets using D6
branes only. 

Our basic guideline throughout this paper is to interpret type IIA
brane configurations in M theory. So we have to identify the type IIA
sixbrane in M theory which was first done in \cite{town}. We consider
M theory on $\mathbb{R}^{10} \times \mathbf 
{S}^1$ which is equivalent to type IIA on $\mathbb{R}^{10}$. The RR $U(1)$
gauge field of type IIA is associated in M theory with shifts along
the $\mathbf{S}^1$. Momentum states in the $\mathbf{S}^1$ direction are
electrically charged with respect to this $U(1)$ and are interpreted
in type IIA as D0 branes. The monopoles of this $U(1)$ correspond to
D6 branes in type IIA. 

The object that is magnetically charged under this $U(1)$ is the
Kaluza-Klein monopole $\mathbb{R}^6\times\tilde Q$, where $\tilde Q$
is a Taub-NUT space. It can be described as
\be
\tilde Q=\left\lbrace (v,y,z)\in\mathbb{C}^3\Big|yz=
\prod_{a=1}^d(v^2-e_a^2)\equiv Q(v^2)\right\rbrace
%y z = \prod_{a=1}^d (v^2 - e_a^2) \equiv Q(v)
\eel{NUT}
Here we have incorporated the fact that the D6 branes
come in pairs located at $\pm e_a$. 
As explained in \cite{witten}, 
asymptotically $y$ and $z$ can be identified with $t$ and $t^{-1}$.
$\tilde Q$ is smooth as long as $e_a\neq e_b\,~\forall\, a\neq b$. 
Otherwise the singularity has to be resolved.

We start with models with a single gauge group and
matter hypermultiplets. To incorporate D6 branes we have to
replace $Q = \mathbb{R}^3 \times \mathbf{S}^1$ by 
$\tilde{Q}$. As before, the type IIA
configuration of D4 and NS fivebranes is described by a complex curve
$\Sigma$, now embedded in $\tilde{Q}$. $\Sigma$ will again be given by an
polynomial equation $F(y,v) = 0$; any dependence on $z$ has been
eliminated using $z = Q(v)/y$. The $C_n$ and $D_n$ configurations are
symmetric under $(v,y,z) \to (-v,y,z)$, the $B_n$ configurations are
symmetric under $(v,y,z) \to (-v,-y,-z)$ due to the asymptotic
relation between $t$, $t^{-1}$, and $y$, $z$. 

With two NS fivebranes,  $F(y,v)$ is quadratic in
$y$. Furthermore we assume that there are no semi-infinite D4 branes
to the left or to the right, except those needed for the orthogonal
gauge groups. In this example we will choose $G = SO(2 r)$ but the
discussion also applies to gauge groups of the $B_r$ and $C_r$
series. Thus $F$ has the form 
\be
A(v) v^2 y^2 + B(v) y + C(v) v^2 = 0,
\eel{aaa}
where $A$, $B$, and $C$ are relatively prime polynomials, which as all 
polynomial appearing here and below, depend on $v$ through $v^2$ only.
The condition that there be no semi-infinite
D4 branes implies that $A$ is constant; we set it to $1$.
Expressing \equ{aaa}  in terms of $z = Q(v^2)/y$ we obtain
\be
C(v) v^2 z^2 + B(v) Q(v) z + Q(v)^2 v^2 = 0.
\ee
The absence of semi-infinite D4 branes implies that $C$ divides $B Q$
and $Q^2$. In particular this means that $Q^2$ is divisible by $C$. So
any zero of $C$ must be a zero of $Q$ and may appear at most
quadratically in $C$. This means we can split $Q$ into three factors:
$Q_0$, $Q_1$, and $Q_2$, whose roots are roots of $C$ of order
$0$, $1$, and $2$, respectively. We will denote the number of zeroes
of the $Q_i$ by $q_i$.
Thus we have
\be
C(v) = f Q_2(v)^2 Q_1(v),
\ee
with $f$ being a non-zero constant.
In addition $B Q = B Q_0 Q_1 Q_2$ has to be divisible by $C$, leading to
\be
B(v) = \tilde{B}(v) Q_2(v).
\ee
for some polynomial $\tilde{B}(v)$.
Now $F$ assumes the form
\be
v^2 y^2 + \tilde{B}(v) Q_2(v)y + v^2 f Q_2(v)^2 Q_1(v) = 0.
\ee
In terms of the coordinate $\tilde{y} = y/Q_2(v)$ this becomes
\be
v^2 \tilde{y}^2 + \tilde{B}(v) \tilde{y} + v^2 f Q_1(v) = 0.
\ee
If $\tilde{B}(v)$ is a polynomial of degree $2 N$ in $v$, this presents
a curve for $SO(2 N)$ with $q_1$ flavors in the fundamental
representation. The zeroes of $Q_1$ correspond to D6 branes between
the two NS fivebranes, the zeroes of $Q_0$ and $Q_2$ correspond to D6
branes to the right and to the left of all NS fivebranes, respectively.
As long as $q_1 \neq 2 N - 2$ we can set $f=1$ by rescaling $v$ and
$\tilde{y}$. 

Finally we want to include D6 branes in the models with chains of NS
fivebranes. The curve $\Sigma$ will now be defined by the zero locus of a
polynomial $F(y,v)$ of the form:
\bea
y^{n+1} + P_1(v) v^2 y^n + P_2(v) y^{n-1} + P_3(v) v^2 y^{n-2} + P_4(v)
y^{n-3} + \ldots  \non
\ldots + P_{n+1}(v) = 0.
\eeal{cccc}
The substitution $y = Q(v)/z$ leads to
\be
P_{n+1} z^{n+1} + v^2 Q P_n z^n + Q^2 P_{n-1} z^{n-1} + \ldots +
Q^{n+1}=0 .
\ee
Absence of semi-infinite D4 branes implies that $Q^m P_{n-m+1}$ is
divisible by $P_{n+1}$. Hence all zeroes of $P_{n+1}$ are zeroes of $Q$
and the multiplicities of the zeroes of $P_{n+1}$ must lie 
between $0$ and $n+1$. Define polynomials
\be
Q_m = \prod_{a = i_m + 1}^{i_{m+1}} (v^2 - e_a^2)\,,\quad
Q=\prod_{m=0}^{n+1}Q_m \,,\quad 0\leq i_0\leq i_1\leq\dots\leq i_n
\ee
such that
\be
P_{n+1} = f \prod_{m=0}^{n+1} Q_m^{n+1 - m}
\ee
and for $1 \leq m \leq n$
\be
P_m = B_m(v) \prod_{j=0}^{m-1} Q_j^{m - j}.
\ee
Via the transformation $y \to y  Q_0$ the curve \equ{cccc}
takes the following 
\bea
y^{n+1} + v^2 B_1(v) y^n + B_2(v) Q_1(v) y^{n-1} + v^2 B_3(v) Q_1(v)^2
Q_2(v) + \ldots \non
\ldots + f (\delta_{-1,(-1)^n}+v^2\delta_{1,(-1)^n})\,
\prod_{j=1}^{n} Q_j^{n+1-j} = 0.
\eea
For $n=1$ this reduces to the case of two NS fivebranes
discussed earlier in this section. For general $n$ we conclude from the 
matter content that $d_s = i_{s+1} - i_s$ is the number of D6 
branes between the $s$-th and the $(s+1)$-th NS fivebrane. 
If the degree of $B_s$ is $2 r_s$ the gauge group is
\be
G = SO(2 r_1) \times Sp(2 r_2) \times SO(2 r_3) \times \ldots.
\ee
The hypermultiplets are in the representations $(2 r_s,
2 r_{s+1})$ plus $d_s$ hypermultiplets in the fundamental
representation of $SO(2 r_s)$($Sp(2 r_s)$) for even(odd) $s$.

\section{Conclusions And Outlook} 

Using the ideas of \cite{witten} we have generalized the
interpretation of the Type IIA brane configuration giving rise to
$N=2$ supersymmetric field theories with and without matter, to all
classical groups. To a large part
this is a straightforward extension of ref. \cite{witten}.
The interesting new aspect is how the orientifold plane, which is
present in the Type IIA picture, manifests itself from the M theory
point of view. Here we have taken the point of view that the
orientifold plane is not directly visible in the Riemann surface which
describes the internal, one-complex-dimensional part of the M5
brane. We only detect it via the presence of additional semi-infinite
D4 branes, which do not add to the spectrum and via the symmetry of
the brane configuration. We do not use the knowledge from Type IIA
theory about the charge assignment of the orientifold plane but try
to understand it from the M-theoretic brane configuration. 
We determined the RR charge of D4 branes, for instance those that stretch to 
$x_6=\pm \infty$ in the $SO(2r)$ case, by observing that they have
non-trivial monodromy around the origin of the $t$ plane, namely,
they are wrapped around the circle in the $x^{10}$ direction. In fact
this maybe a useful tool in general to determine the RR charge. A
fivebrane of M theory, when it wraps the circle $N$ times, implies a
configuration of $N$ D4 branes in the Type IIA description. A similar
situation applies to the NS charge of the strings in Type IIA that
are obtained from wrapping the M theory membrane. Our starting point was
the known curves for the various gauge groups. We find a satisfactory
picture, which can be connected to the Type IIA picture for $B_r$, $C_r$ and
$D_r$ gauge groups. 

The real challenge, in our mind, is the generalization to $N=1$ theories. 
The Type IIA picture has been fully developed for all classical groups.  
The Riemann surface in this case will be embedded in 
$\mathbb{R}^5\times\mathbf{S}^1$.
Presumably it encodes information about the Coulomb branch of the 
$N=1$ theories, but many new insights into $N=1$ theories might be
gained by a better understanding in the framework of M theory.

\section*{Acknowledgements}

We would like to thank D. Ghoshal, S. F\"orste, V. Kaplunovsky
and D. Matalliotakis for instructive discussions.

\end{document}